\documentclass[runningheads]{llncs}
\usepackage[T1]{fontenc}
\usepackage{graphicx}
\usepackage{color}
\usepackage{hyperref}
\usepackage{mathtools}
\usepackage{algorithm}
\usepackage[noend]{algpseudocode}
\usepackage{tabularx}
\usepackage{multirow}
\usepackage{arydshln}

\newcommand{\RICSPEEDUP}{3.61}
\newcommand{\RICMORE}{23}

\begin{document}
\title{Deeply Optimizing the SAT Solver for the IC3 Algorithm}

\author{
    Yuheng Su\inst{1,2} \and
    Qiusong Yang\inst{1,3}\thanks{Qiusong Yang is the corresponding author.} \and
    Yiwei Ci\inst{1} \and
    Yingcheng Li\inst{1,2} \and
    Tianjun Bu\inst{1,2}\and
    Ziyu Huang\inst{4}}
\authorrunning{Y. Su et al.}
\institute{ 
    Institute of Software, Chinese Academy of Sciences \and
    University of Chinese Academy of Sciences \and
    Advanced Computing and Intelligence Engineering, Wuxi, China \and
    Beijing Forestry University \\
    \email{gipsyh.icu@gmail.com} \\
    \email{\{qiusong,yiwei\}@iscas.ac.cn} \\
    \email{\{liyingcheng18,butianjun24\}@mails.ucas.ac.cn} \\
    \email{fyy0007@bjfu.edu.cn}
}

\maketitle
\begin{abstract}
The IC3 algorithm, also known as PDR, is a SAT-based model checking algorithm that has significantly influenced the field in recent years due to its efficiency, scalability, and completeness. It utilizes SAT solvers to solve a series of SAT queries associated with relative induction. In this paper, we introduce several optimizations for the SAT solver in IC3 based on our observations of the unique characteristics of these SAT queries. By observing that SAT queries do not necessarily require decisions on all variables, we compute a subset of variables that need to be decided before each solving process while ensuring that the result remains unaffected. Additionally, noting that the overhead of binary heap operations in VSIDS is non-negligible, we replace the binary heap with buckets to achieve constant-time operations. Furthermore, we support temporary clauses without the need to allocate a new activation variable for each solving process, thereby eliminating the need to reset solvers. We developed a novel lightweight CDCL SAT solver, GipSAT, which integrates these optimizations. A comprehensive evaluation highlights the performance improvements achieved by GipSAT. Specifically, the GipSAT-based IC3 demonstrates an average speedup of $\RICSPEEDUP$ times in solving time compared to the IC3 implementation based on MiniSat.

\end{abstract}

\section{Introduction}
Model checking \cite{ModelChecking,HandbookMC} is a powerful formal verification technique widely used in system design. Given a transition system and a property describing the desired system behavior, it can efficiently and automatically detect bugs or prove system correctness.

Recent significant advancements in SAT solvers, particularly the introduction of the Conflict Driven Clause Learning algorithm (CDCL) \cite{CDCL}, have led to breakthroughs in model checking through SAT-based algorithms. Among these, IC3 \cite{IC3} (also known as PDR \cite{PDR}) stands out as a highly influential SAT-based model checking algorithm, widely used for hardware verification. IC3 efficiently searches for inductive invariants without requiring model unrolling. Compared to Bounded Model Checking (BMC) \cite{BMC}, IC3 offers completeness, and it demonstrates superior scalability over Interpolation-based Model Checking (IMC) \cite{IMC} and K-Induction \cite{KINDUCTION}. As a result, IC3 is widely regarded as the state-of-the-art algorithm and serves as the core engine in many efficient model checkers \cite{NUXMV,AVR,PONO}.

IC3 works by constructing SAT problems regarding the overapproximate reachable states of the model being verified. With the help of a SAT solver, solving these problems yields answers that facilitate the verification process. Therefore, currently, almost all implementations of IC3 essentially have a two-layer structure: the top layer is the IC3 algorithm layer, which drives the verification process by posing a series of SAT queries, and the bottom layer is the SAT solver layer, which handles the queries posed by the algorithm layer.

IC3 has been successfully applied to verify some industrial designs \cite{IC3IndustrialDesign}. However, it still faces challenges due to the problem of state space explosion. Therefore, enhancing the performance and scalability of the IC3 algorithm to efficiently verify larger-scale models continues to be an important research direction. Most studies aiming to improve the IC3 algorithm focus on optimizing at the algorithm layer by proposing various variants or enhancing the efficiency of intermediate steps \cite{CTG,PDRPROGRESS,PredictingLemma,CAV23}. Only a very few studies focus on the SAT solver layer: \cite{IC3SolverManagement} compared the performance of different SAT solver management strategies at the algorithm layer. \cite{TemporaryClause} proposed a simple and easily implementable approach that enables the addition of a temporary clause without the need for activation literals, eliminating the need for IC3 to reset the SAT solver.

In this paper, we propose several optimizations for the SAT solver used in IC3, based on our analysis of the unique characteristics of SAT queries. We introduce a lightweight SAT solver, \textbf{GipSAT}, which incorporates these optimizations to enhance its efficiency in processing SAT queries within IC3. The key contributions of this paper are summarized as follows:
\begin{itemize}
    \item We analyzed the characteristics of SAT queries in IC3 and observed that it is not necessary to make decisions on all variables. Additionally, the overhead of binary heap operations in VSIDS is non-negligible.
    \item We propose a method that, before each solving step, analyzes a subset of variables that need to be decided, ensuring that the result remains unaffected, thus reducing the number of decisions.
    \item We introduce a novel data structure that replaces the binary heap, enabling VSIDS operations to be performed in constant time.
    \item We propose a method that supports temporary clauses by reusing the same activation variable instead of allocating a new one before each solving process, thus eliminating the need for SAT solver resetting.
    \item We introduce a novel lightweight SAT solver, GipSAT, which incorporates these optimizations. We provide a detailed interface for GipSAT.
    \item We conduct a comprehensive performance evaluation of GipSAT. The experimental results demonstrate that GipSAT significantly enhances the efficiency and scalability of the IC3 algorithm.
\end{itemize}

\section{Preliminaries}
\subsection{Basics and Notations}
We use notations such as $x, y$ for Boolean variables, and $X, Y$ for sets of Boolean variables. The terms $x$ and $\lnot x$ are referred to as literals. A cube is a conjunction of literals, while a clause is a disjunction of literals. A Boolean formula in Conjunctive Normal Form (CNF) is a conjunction of clauses. It is often convenient to treat a clause or a cube as a set of literals, and a CNF as a set of clauses. For instance, given a CNF formula $F$, a clause $c$, and a literal $l$, we write $l \in c$ to indicate that $l$ occurs in $c$, and $c\in F$ to indicate that $c$ occurs in $F$. A formula $F$ implies another formula $G$, if every satisfying assignment of $F$ satisfies $G$, denoted as $F \Rightarrow G$. We use $\mathcal{V}(F)$ to represent the set of all variables appeared in formula $F$.

A transition system, denoted as $S$, can be defined as a tuple $\langle X, Y, I, T\rangle$. Here, $X$ and $X'$ represent the sets of state variables for the current state and the next state, respectively, while $Y$ represents the set of input variables. The Boolean formula $I(X)$ represents the initial states, and $T(X, Y, X')$ describes the transition relation. State $s_2$ is a successor of state $s_1$ if and only if $(s_1, s_2') \Rightarrow T$. A safety property $P(X)$ is a Boolean formula over $X$. A system $S$ satisfies $P$ if and only if all reachable states of $S$ satisfy $P$.

Without loss of generality, circuits are commonly represented in the form of And-Inverter Graph (AIG) \cite{AIG}. An AIG is a directed acyclic graph, which includes primary inputs/outputs and two input and-nodes with optional inverter marks on the fanin edges. The Cone of Influence (COI) of a node is the set of all nodes that could potentially influence its value, which can be obtained by recursively traversing its fanins. When $T$ is derived from an AIG, $X$ and $Y$ correspond to the primary inputs, while $X'$ corresponds to the primary outputs. It exhibits the functional characteristic where each next state variable is assigned by a function of current states and inputs, $x_i' \leftrightarrow f_i(X, Y)$. Therefore, $T$ is a conjunction of all assignment functions, $\bigwedge x_{i}' \leftrightarrow f_i(X, Y)$. The transformation of an AIG to CNF typically involves Tseitin encoding \cite{Tseitin}, where each node in the AIG is mapped to a variable in the CNF.

\subsection{Incremental CDCL SAT Solver}
Conflict Driven Clause Learning (CDCL) \cite{CDCL} is a powerful and widely used algorithm that employs conflict analysis and clause learning techniques to efficiently solve SAT problems. Modern CDCL SAT solvers typically rely on the Variable State Independent Decaying Sum (VSIDS) branching heuristic \cite{VSIDS}. It calculates a score for each variable and selects the variable with the highest score during decision. This is typically maintained by a binary heap \cite{BinaryHeap}. An incremental SAT solver efficiently solves a series of related formulas and typically provides the following interfaces:
\begin{itemize}
    \item \texttt{add\_clause($clause$)}: adds the $clause$ to the solver.
    \item \texttt{solve($assumption$)}: checks the satisfiability under the given $assumption$.
    \item \texttt{get\_model()}: retrieves the variable assignments from the previous SAT call.
    \item \texttt{unsat\_core()}: retrieves the unsatisfiable core from the assumptions of the previous UNSAT call.
\end{itemize}

\subsection{IC3 Algorithm}
\begin{algorithm}
\caption{Overview of IC3}
\label{alg:ic3}
\begin{algorithmic}[1]
\Function{$get\_predecessor$()}{}
    \State $model \coloneqq get\_model()$
    \State \Return $\{l \in model \mid var(l) \in X\}$
\EndFunction
\\
\Function{$generalize$}{cube $b$, level $i$}
    \For{each $l \in b$}
        \State $cand \coloneqq b \setminus \{l\}$
        \If{$I \Rightarrow \lnot cand$ and $\lnot sat(F_{i-1}\land \lnot cand \land T \land cand')$} \Comment{$Q_{gen}$}
            \State $b \coloneqq unsat\_core()$
        \EndIf
    \EndFor
    \State \Return $b$
\EndFunction
\\
\Function{$block$}{cube $c$, level $i$}
    \If{$i = 0$}
        \State \Return $false$
    \EndIf
    \While{$sat(F_{i-1}\land \lnot c \land T \land c')$} \Comment{$Q_{block}$}
        \State $p \coloneqq get\_predecessor()$
        \If{$\lnot block(p, i-1)$}
            \State \Return $false$
        \EndIf
    \EndWhile
    \State $c \coloneqq unsat\_core()$
    \State $lemma \coloneqq \lnot generalize(c, i)$
    \For{$1\leq j \leq i$} \label{ic3:addlemma}
        \State $F_j \coloneqq F_j \cup \{lemma\}$
    \EndFor
    \State \Return $true$
\EndFunction
\\
\Function{$propagate$}{level $k$} \label{ic3:propagate}
    \For{$1 \leq i < k$}
        \For{each $c \in F_i \setminus F_{i+1}$} 
            \If{$\lnot sat(F_i \land T \land \lnot c')$} \Comment{$Q_{push}$}
                \State $F_{i+1} \coloneqq F_{i+1} \cup \{c\}$
            \EndIf
        \EndFor
        \If{$F_i = F_{i+1}$}
            \State \Return $true$
        \EndIf
    \EndFor
    \State \Return $false$
\EndFunction
\\
\Procedure{IC3}{$I,T,P$}
    \State $F_0 \coloneqq I,k \coloneqq 1,F_k \coloneqq \top$
    \While{$true$}
        \While{$sat(F_k\land \lnot P)$} \Comment{$Q_{bad}$}
            \State $c \coloneqq get\_predecessor()$
            \If{$\lnot block(c, k)$}
                \State \Return $unsafe$
            \EndIf
        \EndWhile
        \State $k \coloneqq k + 1,F_k \coloneqq \top$
        \If{$propagate(k)$}
            \State \Return $safe$
        \EndIf
    \EndWhile
\EndProcedure
\end{algorithmic}
\end{algorithm}

IC3 is a SAT-based model checking algorithm that does not require unrolling the system \cite{IC3}. It attempts to prove that $S$ satisfies $P$ by finding an inductive invariant $INV(X)$ such that:
\begin{itemize}
\item $I(X) \Rightarrow INV(X)$
\item $INV(X) \land T(X,Y,X') \Rightarrow INV(X')$
\item $INV(X) \Rightarrow P(X)$
\end{itemize}

To achieve this objective, it maintains a monotone CNF sequence $F_0, F_1 \ldots F_k$. Each \emph{frame} $F_i$ is a Boolean formula over $X$, which represents an over-approximation of the states reachable within $i$ transition steps. Each clause $c$ in $F_i$ is called \emph{lemma} and the index $i$ is called \emph{level}. IC3 maintains the following invariant: 

\begin{itemize}
\item $F_0 = I$
\item $F_{i+1} \subseteq F_i$
\item $F_i \land T \Rightarrow F_{i+1}'$
\item for all $i < k, F_i \Rightarrow P$
\end{itemize}

A lemma $c$ is said to be \emph{inductive relative} to $F_i$ if, starting from states that satisfy both $F_i$ and $c$, all states reached in a single transition satisfy $c$. This condition can be expressed as a SAT query $sat(F_i \land c \land T \land \lnot c')$. If this query is satisfiable, it indicates that $c$ is not inductive relative to $F_i$, as we can find a counterexample where a state satisfying $F_i \land c$ and transitions to a state that does not satisfy $c$. If lemma $c$ is inductive relative to $F_i$, it can also be said that the cube $\lnot c$ is blocked in $F_{i+1}$.

Algorithm \ref{alg:ic3} provides an overview of the IC3 algorithm. This algorithm incrementally constructs frames by iteratively performing the blocking phase and the propagation phase. During the blocking phase, the IC3 algorithm focuses on making $F_k \Rightarrow P$. It iteratively get a cube $c$ such that $c \Rightarrow \lnot P$, and block it recursively. This process involves attempting to block the cube's predecessors if it cannot be blocked directly. It continues until the initial states cannot be blocked, indicating that $\lnot P$ can be reached from the initial states in $k$ transitions, which violates the property. In cases where a cube can be confirmed as blocked, IC3 proceeds to expand the set of blocked states through a process called generalization. This involves dropping literals and ensuring that the resulting clause remains relatively inductive, with the objective of obtaining a minimal inductive clause. The propagation phase tries to push lemmas to the top frame. If a lemma $c$ in $F_i \setminus F_{i+1}$ is also inductive relative to $F_i$, then push it into $F_{i+1}$. During this process, if two consecutive frames become identical ($F_i = F_{i+1}$), then the inductive invariant is found and the safety of this model can be proved.

The IC3 algorithm involves four different types of SAT queries, as denoted in Algorithm \ref{alg:ic3}. Since $c \in F_i$, the query $Q_{push}$ can be expressed as $sat(F_i \land c \land T \land \lnot c')$. Consequently, $Q_{gen}$, $Q_{block}$, and $Q_{push}$ are essentially the same and can be uniformly represented as $sat(F_i \land c \land T \land \lnot c')$, where $c$ is a clause and $\lnot c'$ is a cube. This representation captures the inductiveness of $c$ relative to $F_i$, which we denote as $Q_{relind}$.

\section{Motivations}
\begin{table}
\caption{The average solving time, number of calls, and average length of $c$ in $Q_{relind}$ were calculated for various SAT queries, yielding an overall average across all cases.}
\label{tab:SATQueries}
\small
\centering 
\begin{tabular}{c c @{\hspace{2em}} c @{\hspace{2em}} c}
\hline
& Avg. solving time(ms) & \textbf{\#}Calls & Avg. $c$ length \\
\hline
$Q_{gen}$ & 1.099 & 110794.41 & 6.775 \\
$Q_{block}$ & 0.600 & 12814.11 & 104.202 \\
$Q_{push}$ & 0.712 & 87831.33 & 4.643 \\
$Q_{bad}$ & 1.655 & 247.48 & - \\
\hline
\end{tabular}
\end{table}

We analyzed the characteristics of SAT queries in Minisat-based rIC3 \cite{RIC3}. The analysis was conducted on the complete benchmarks of the 2020 and 2024 Hardware Model Checking Competition (HWMCC). For a detailed description of the experimental setup, please refer to Section \ref{Sec:Evaluation}. The results are presented in Table \ref{tab:SATQueries}. It can be observed that the SAT queries generated by IC3 exhibit unique characteristics: they are solved quickly, with an average solving time of less than 1ms, but the number of queries is very large. Additionally, we have the following observations:

\begin{itemize}
\item \textbf{Observation 1}: $Q_{relind}$: $sat(F_{i}\land c \land T \land \lnot c')$ represents whether there exists a predecessor of any state in $\lnot c$ (where $\lnot c$ is a cube, representing a set of states) that satisfies $F_{i}$ but is not in $\lnot c$. We have observed that solving $Q_{relind}$ does not require considering assignments for all variables. Fig. \ref{fig:COI} shows an AIG model. The values of variables in $c'$ are determined solely by $COI(c')$, as illustrated by the shaded region. The values of variables outside the shaded region do not affect the result. From the last column of Table \ref{tab:SATQueries}, it can be observed that the $c$ in $Q_{gen}$ and $Q_{push}$ is typically very small in size. Consequently, the size of $COI(c')$ is likely to be small compared to the total number of variables. We also calculated the average percentage of variables in $COI(c')$ for all $Q_{relind}$ relative to the total number of variables, as shown in Fig. \ref{fig:Domain}. It can be observed that, in most cases, the number of variables that need to be determined is only around 20\% of the total number of variables. We may be able to use this information to reduce the number of variables that need to be assigned in each solving iteration. However, regular SAT solvers can only obtain the CNF representation of the AIG and cannot capture the dependency relationships between variables.

\begin{figure}[!t]
    \centering
    \includegraphics[width=160pt]{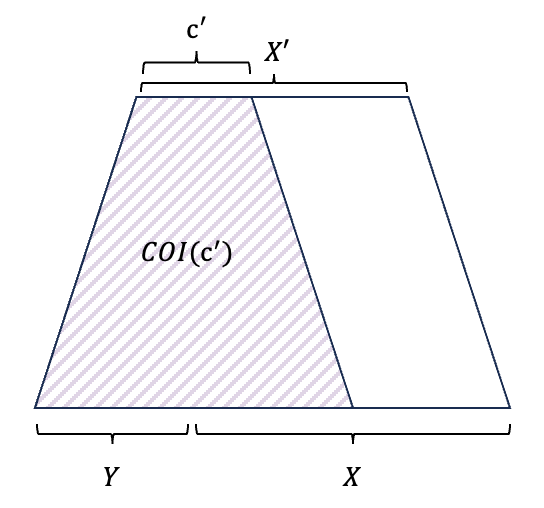}
    \caption{The diagram of an AIG model. The entire trapezoid represents the transition relations $T$, and the shaded region represents the subset of relations required by $c'$.}
    \label{fig:COI}
\end{figure}

\begin{figure}[!t]
    \centering
    \includegraphics[width=0.7\textwidth]{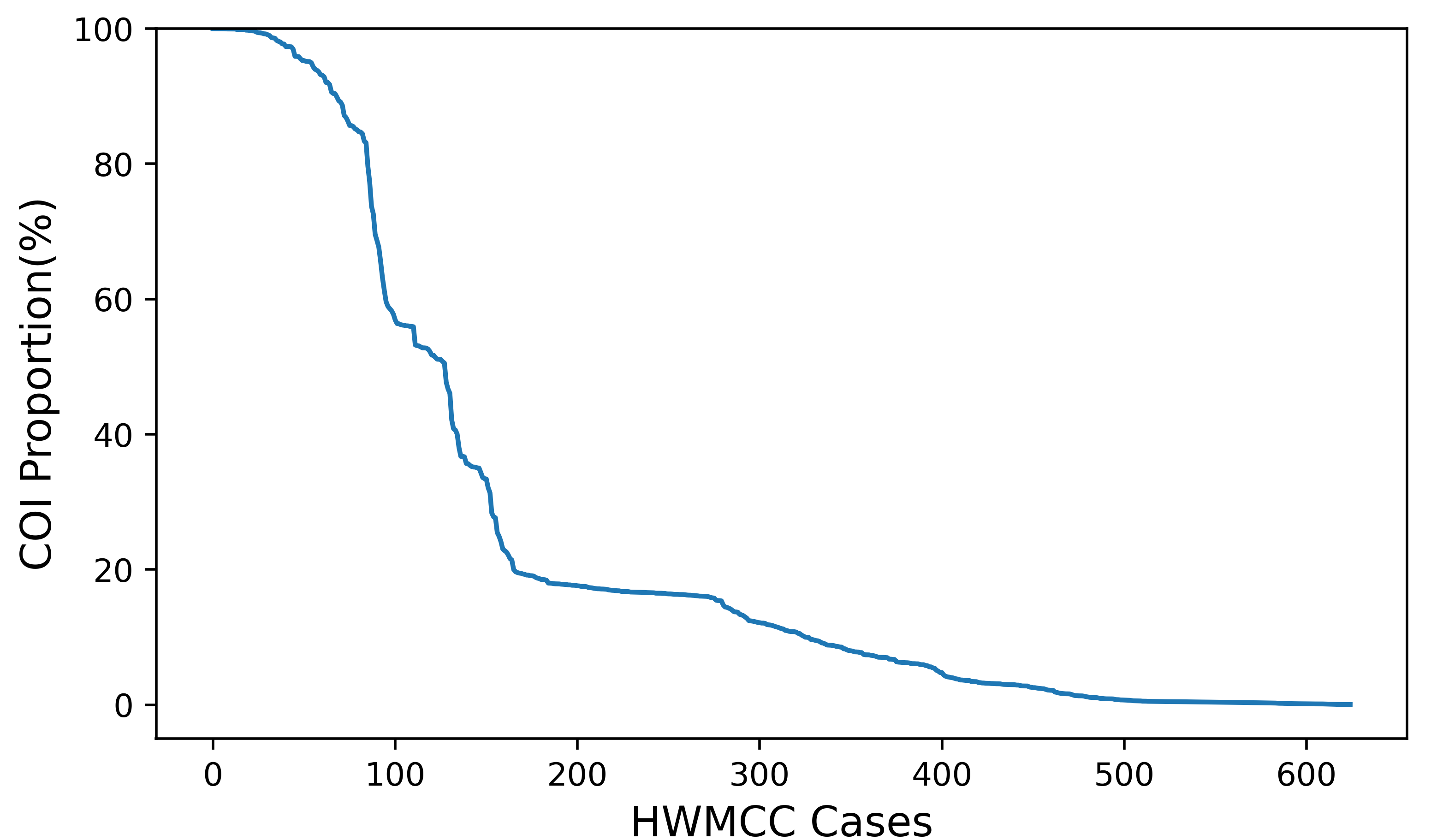}
    \caption{The number of cases (x-axis) in which the average percentage of variables in $COI(c')$ relative to the total number of variables in $Q_{relind}$ exceeds a given value (y-axis).}
    \label{fig:Domain}
\end{figure}

\begin{figure}
    \centering
    \includegraphics[width=0.7\textwidth]{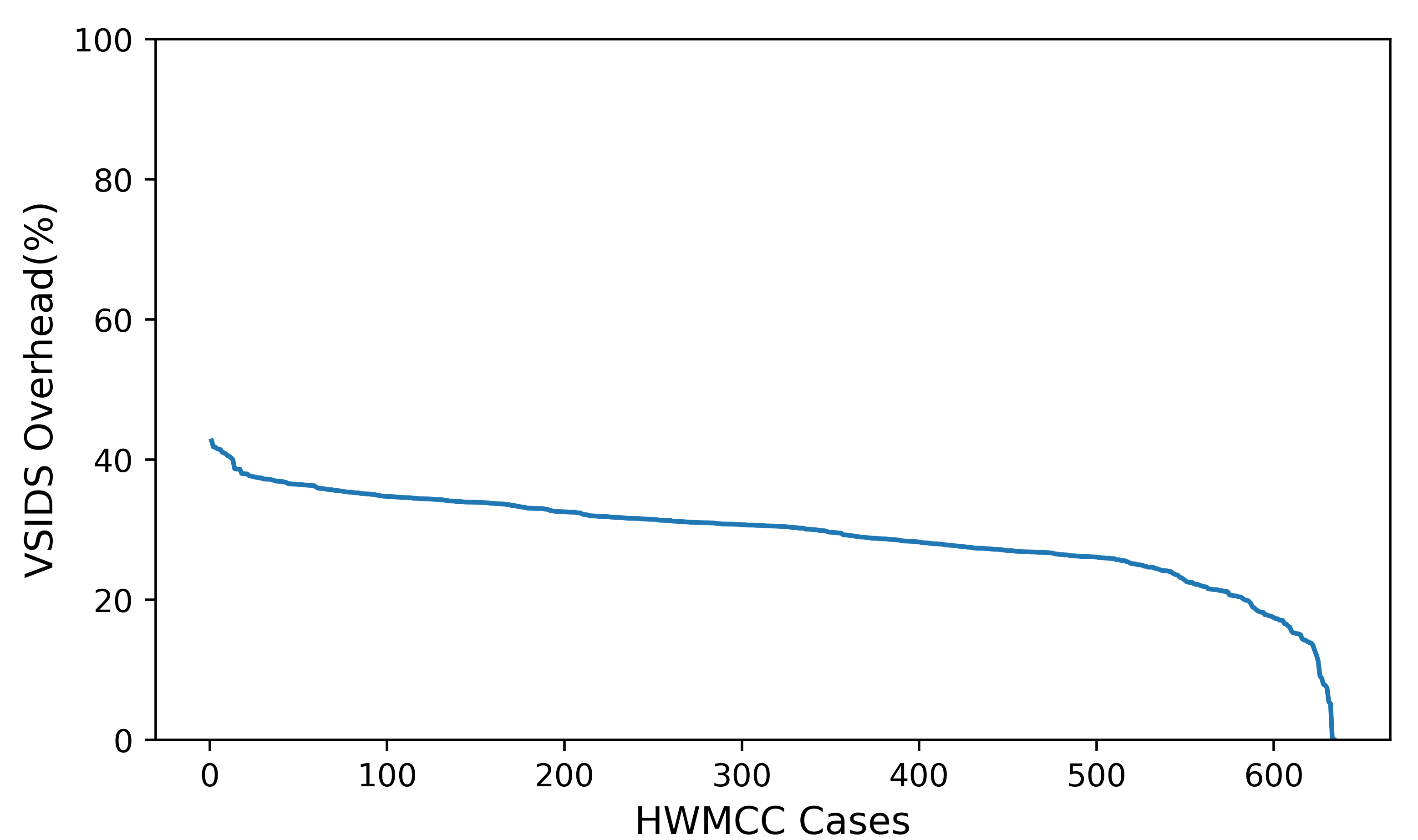}
    \caption{The number of cases (x-axis) in which the overhead of VSIDS operations exceeds a given value (y-axis).}
    \label{fig:Vsids}
\end{figure}

    \item \textbf{Observation 2}: SAT solvers utilize VSIDS (Variable State Independent Decaying Sum) to guide variable decisions. Typically, VSIDS employs a binary heap, where variables with higher scores are prioritized and popped out first. We also measured the overhead of VSIDS operations, as illustrated in Fig. \ref{fig:Vsids}. The results show that the overhead of VSIDS operations is non-negligible, with the majority of cases exhibiting an overhead of around 30\%. This phenomenon may be attributed to the simplicity of the queries posed by IC3, which involve fewer conflicts during solving and may only require variable assignments in decision and BCP (Boolean Constraint Propagation). Due to the logarithmic time complexity of binary heap operations, compared to the constant time required for assignments, the overhead becomes significant.

    \item \textbf{Observation 3}: It can be observed that $Q_{relind}$ requires the addition of a temporary clause $c$ to the SAT solver, which only takes effect in the next solving iteration. However, most state-of-the-art SAT solvers do not support the direct removal of a clause, as removing a clause from the formula would necessitate removing every learned clause derived from it. One common approach to address this issue involves the use of activation variables \cite{ACTVAR}, which most IC3 implementations adopt. Let $a$ be an activation literal, which is a free variable that does not occur within the original formula. By incorporating $c \lor a$ into the formula and assuming $\lnot a$, we effectively achieve the same result as directly adding $c$. To remove $c$, we can simply add the clause consisting of the single literal $a$ to the formula. However, solving $Q_{relind}$ each time requires creating a new activation variable. As the number of activation variables increases, their presence can negatively impact the performance of the SAT solver \cite{IC3SolverManagement,TemporaryClause}. A common solution to this problem is to periodically reset the SAT solver with a fresh instance, though this approach has the drawback of clearing all learned clauses and heuristic scores. Our goal is to support temporary clauses while avoiding the need for such resets.

\end{itemize}

We can potentially leverage these distinctive characteristics to optimize the SAT solver for the IC3 algorithm to improve its performance.

\section{Optimizations}
\subsection{Decide and Propagate in Domain}
\definecolor{deepblue}{RGB}{0,0,148}
Observation 1 shows that it is unnecessary to consider all variables when solving $Q_{relind}$. We analyze which variables need to be considered through the following theorems. Considering the formula $Q_{relind}$: $sat(F_i \land c \land T \land \lnot c')$, we can analyze the subset of variables that need to be taken into account without affecting the satisfiability. To determine the satisfiability of $F_i \land c$, we only need to consider the variables that appear in this formula, which are $\mathcal{V}(F_i) \cup \mathcal{V}(c)$. Similarly, for the satisfiability of $T \land \lnot c'$, the values of the variables in $\lnot c'$ are solely determined by the $COI(c')$. Therefore only the variables in $COI(c')$ need to be considered. As a result, by considering only the variables in $\mathcal{V}(F_i) \cup \mathcal{V}(c) \cup COI(c')$, we can determine the satisfiability of the formula $Q_{relind}$.

We refer to the set of variables that need to be considered during solving as the \emph{domain}. For solving $Q_{relind}$, it is sufficient to use $\mathcal{V}(F_i) \cup \mathcal{V}(c) \cup COI(c')$ as the domain, instead of all variables. This significantly reduces the number of variables that the SAT solver needs to decide and assign during BCP. 

We use a hash table to record whether each variable belongs to the domain, allowing constant-time queries for domain information. Before each solving process, we need to compute the domain relevant to that instance. For $\mathcal{V}(F_i)$, the variables are permanently stored in the hash table, and new domain variables are added whenever a new lemma is introduced. For $\mathcal{V}(c)$, we simply extract all variables that appear in clause $c$. Finally, $COI(c')$ is computed by recursively traversing the fanins of $c'$. After computing $\mathcal{V}(c)$ and $COI(c')$, we also temporarily record them in the domain hash table, but they are removed after each solving process.

After computing the domain for the current solving instance, we push all domain variables into the VSIDS decision-required set to ensure they are considered for decisions. During BCP, if a clause contains a variable that is not in the current domain, it is considered irrelevant to the current solving process. Since out-of-domain variables are not eligible for assignment, any clause involving them cannot contribute to further propagation. When BCP attempts to find a new unassigned variable to serve as the watch literal, if that variable lies outside the domain, we simply ignore the clause and proceed to the next one, without performing any variable assignments based on that clause. This approach guarantees that out-of-domain variables are never assigned during BCP. Furthermore, by restricting assignments to domain variables only, we eliminate propagation involving irrelevant variables, thereby reducing BCP overhead.

It can also be noticed that the difference between $c'$ in every two adjacent $Q_{gen}$ is only one literal. This suggests that calculating $COI(c')$ for every $Q_{gen}$ is unnecessary. Instead, We perform this calculation initially during generalization and upon successful drops, as depicted in Algorithm \ref{alg:coigeneralize}. We recalculate it upon successful drops because sometimes the unsat core can drop multiple variables at once compared to before.

\begin{algorithm}
\caption{Generalize with Domain}
\label{alg:coigeneralize}
\begin{algorithmic}[1]
\Function{$generalize$}{cube $b$, level $i$}
    \State \textcolor{deepblue} { $set\_domain(\mathcal{V}(F_{i-1} \land \lnot b) \cup COI(b'))$ }
    \For{each $l \in b$}
        \State $cand \coloneqq b \setminus \{l\}$
        \If{$I \Rightarrow \lnot cand$ and $\lnot sat(F_{i-1}\land \lnot cand \land T \land cand')$} \Comment{$Q_{gen}$}
            \State $b \coloneqq unsat\_core()$
            \State \textcolor{deepblue} { $set\_domain(\mathcal{V}(F_{i-1} \land \lnot b) \cup COI(b'))$ }
        \EndIf
    \EndFor
    \State \textcolor{deepblue} { $unset\_domain()$ }
    \State \Return $b$
\EndFunction
\end{algorithmic}
\end{algorithm}

\subsection{VSIDS by Bucket}

\begin{algorithm}[!t]
\caption{VSIDS by Bucket}
\label{alg:bucket}
\begin{algorithmic}[1]
\State $IntervalHeap\  iheaps[NumBucket]$ \Comment{Interval heap for each bucket}
\State $Map\  var\_bucket$ \Comment{Hash table maps variables to buckets}
\State $Queue\  queue[NumBucket]$ \Comment{Variable queue for each bucket}
\\
\Function{$update$}{Var $v$}
    \State $b \coloneqq var\_bucket[v]$
    \While{$b > 0$}
        \If{$iheaps[b].max().score > iheaps[b-1].min().score$}
            \State $vmax \coloneqq iheaps[b].pop\_max()$
            \State $vmin \coloneqq iheaps[b-1].pop\_min()$
            \State $iheaps[b].push(vmin)$
            \State $iheaps[b-1].push(vmax)$
            \State $var\_bucket[vmin] = b$
            \State $var\_bucket[vmax] = b - 1$
        \Else
            \State \textbf{break}
        \EndIf
        \State $b = b - 1$
    \EndWhile
\EndFunction
\\
\Function{$push$}{Var $v$}
    \State $bucket \coloneqq var\_bucket[v]$
    \State $queue[bucket].push\_back(v)$
    \State $head = min(head, bucket)$
\EndFunction
\\
\Function{$pop$}{ }
    \While{$head < NumBucket$}
        \If{$queue[head].size() > 0$}
            \State \Return $queue[head].pop\_front()$
        \EndIf
        \State $head = head + 1$
    \EndWhile
    \State \Return $None$
\EndFunction
\end{algorithmic}
\end{algorithm}

Observation 2 highlights that the overhead of VSIDS is non-negligible, and reducing this overhead could potentially enhance overall performance. Inspired by bucket sorting, we introduce a novel data structure for VSIDS, that enables both push and pop operations to be performed in constant time. We predefine a fixed number of buckets (defaulting to 15) and assign each variable to a specific bucket. This design ensures that variables in lower-numbered buckets have higher VSIDS scores than those in higher-numbered buckets.

To maintain this guarantee amid dynamic changes in variable scores, we utilize an interval heap for each bucket. The interval heap is a double-ended priority queue that can efficiently pop either the maximum or minimum value in logarithmic time \cite{IntervalHeap}. Each bucket’s interval heap stores all the variables belonging to that bucket, using their scores as the sorting criterion. When the score of variable $v$ increases, the $update$ function in Algorithm \ref{alg:bucket} is invoked. This function first locates the bucket $b$ where $v$ resides. Then, it checks whether the maximum score in bucket $b$ is greater than the minimum score in bucket $b-1$. If so, it pops the variable with the maximum score from bucket $b$ and pushes it into bucket $b-1$. Next, it pops the variable with the minimum score from bucket $b-1$ and pushes it into bucket $b$. This process is repeated on bucket $b-1$ to ensure the monotonicity of scores across buckets.

The interval heap is used to maintain the bucket corresponding to each variable, with changes occurring only when the score is updated. Additionally, we create a queue for each bucket to store the variables waiting to be decided within that bucket. The variables in the queue are unordered. During variable decision, the $pop$ function in Algorithm \ref{alg:bucket} selects a variable from the non-empty queue of the smallest bucket. During backtracking, the $push$ function places unassigned variables back into their respective queues based on their buckets.

This approach retains logarithmic time complexity for score updates while reducing the time complexity of the push and pop operations in VSIDS from logarithmic to constant time. However, this improvement comes at the cost of reduced accuracy in VSIDS, as the variable selected each time is from the bucket with the highest score but not necessarily the highest-scoring variable within that bucket. Nonetheless, since the queries are relatively simple, very high accuracy may not be critical.

\subsection{Without Resetting}
\label{Sec:WithoutResetting}
Observation 3 indicates that the SAT solver requires periodic resets due to the activation variables. However, this process also results in the loss of all learned clauses and heuristic scores. The key to supporting temporary clauses is to identify and remove all learned clauses that are directly or indirectly derived from temporary clauses after each solving process. We establish the following theorem:

\begin{theorem}
\label{theorem:learntclause}
To solve $Q_{relind}$ with a temporary clause $\mathbf{c}$ by utilizing the activation variable $\mathbf{a}$, we consider the formula: $
\text{sat}(F_{i} \land (c \lor a) \land T \land \lnot c' \land \lnot a)$, where $\lnot c' \land \lnot a$ serves as the temporary assumption. If a learned clause $\mathbf{lc}$ is derived from $c \lor a$, then $\mathbf{lc}$ must contain the literal $\mathbf{a}$.
\end{theorem}

\begin{proof}
In a CDCL solver, every learned clause is the result of a sequence of resolution steps \cite{HandbookSAT,SATMC}. We prove this theorem by mathematical induction. Initially, the only clause containing $a$ is $c \lor a$, and we assume that no clause contains the literal $\lnot a$. Any learned clause derived from $c \lor a$ is either a direct or indirect resolvent of $c \lor a$. Since we assume that no clause contains $\lnot a$, $a$ cannot be used as a pivot in resolution. Consequently, every resolution step involving $c \lor a$ must preserve the literal $a$, implying that all learned clauses derived from $c \lor a$ must contain $a$. Furthermore, clauses not derived from $c \lor a$ cannot contain either $a$ or $\lnot a$, ensuring that our assumption remains valid throughout the process.
\end{proof}

Theorem \ref{theorem:learntclause} states that learned clauses derived from the temporary clause contain the activation variable. Therefore, if all learned clauses containing the activation variable are identified and removed after each solving process, the temporary clause can also be safely deleted.

We use only one activation variable to support the temporary clause. A fixed variable is allocated in advance and serves as the activation variable for every query. During conflict clause analysis, if the learned clause contains this variable, it is recorded in a special vector. After each solving process, all learned clauses stored in this vector are removed from the clause database, along with the temporary clause itself. As a result, both the temporary clause and all its derived learned clauses are eliminated, ensuring that the activation variable does not appear in any remaining clauses. This frees the variable, making it available for reuse in the next solving process. Since this approach requires only a single additional variable, it eliminates the need to reset the SAT solver.

\section{Implementation}
Instead of implementing these optimizations on an existing SAT solver, we developed a new lightweight CDCL-based SAT solver from scratch, called GipSAT. It is implemented in Rust, a modern programming language designed to offer both performance and safety.

In many current implementations of IC3 \cite{ABC,IC3ref,RIC3}, a separate SAT solver is created for each frame. GipSAT is designed based on this assumption, as each GipSAT instance corresponds to a specific frame. Since GipSAT is specifically optimized for the IC3 algorithm, it requires more information than other regular SAT solvers, such as the transition relation and awareness of which clauses belong to a frame for computing and maintaining the domain. Consequently, its interface differs from that of regular SAT solvers. The interfaces provided by GipSAT are as follows:

\begin{itemize}
    \item \texttt{new($model$)}: Creates a new instance of GipSAT based on the provided transition system $model$. The transition relation $T$ of $model$ is represented in CNF format, with variable dependency information derived from AIG to compute the COI.

    \item \texttt{add\_lemma($lemma$)}: Adds the given \( \textit{lemma} \) to the solver. GipSAT maintains the lemmas of each frame. The \texttt{add\_clause} method has been replaced with this, as users only need to add lemmas after creating a GipSAT instance. A lemma is a clause that involves variables in $X$ and has no dependent variables. Adding a clause with dependent variables could affect the correctness of domain maintenance. Therefore, only the \texttt{add\_lemma} interface is provided.

    \item \texttt{solve($assumption,constraint,droot$)}: Checks the satisfiability under the specified $assumption$ and $constraint$ within the domain $\text{COI}(droot) \cup \mathcal{V}(F)$, where $F$ represents the lemmas maintained by GipSAT. Users should ensure that $\mathcal{V}(assumption) \cup \mathcal{V}(constraint) \subseteq \text{COI}(droot) \cup \mathcal{V}(F)$.

    \item \texttt{unsat\_core()}: Retrieves the UNSAT core from the previous UNSAT call.

    \item \texttt{get\_model()}: Retrieves the variable assignments from the previous SAT call.

    \item \texttt{set\_domain($droot$)}: Configures GipSAT to use $\mathcal{V}(F) \cup \text{COI}(droot)$ as the domain for subsequent solving processes, bypassing the need to recompute the domain before each solving step.

    \item \texttt{unset\_domain()}: Resets the previously configured domain setting.
\end{itemize}

Meanwhile, GipSAT also provides encapsulated interfaces for the IC3 algorithm by leveraging the aforementioned functionalities. These interfaces include:

\begin{itemize}
    \item \texttt{relind($c$)}: Checks the satisfiability of $Q_{relind}$, which is functionally equivalent to \texttt{solve($\lnot c',c,\mathcal{V}(c)\cup \mathcal{V}(c')$)}.

    \item \texttt{inductive\_core()}: The inductive core is the UNSAT core produced by the previous \texttt{relind($c$)}, and an extra condition ensures that it does not overlap with the initial state $I$.

    \item \texttt{has\_bad()}: Checks the satisfiability of $Q_{bad}$, which is functionally equivalent to \texttt{solve($P,true,\mathcal{V}(P)$)}.
\end{itemize}

\section{Evaluation}
\label{Sec:Evaluation}

\subsection{Experiment Setup}
We integrated GipSAT into rIC3 \cite{RIC3}, an implementation of the IC3 algorithm in Rust, which boasts competitive performance. We use the version based on Minisat. For comparison, we evaluated the performance of rIC3 with some state-of-the-art SAT solvers:

\begin{itemize}
    \item \textbf{Minisat 2.2.0} \cite{Minisat}: Minisat is a popular SAT solver that serves as the underlying framework for many other SAT solvers.
    \item \textbf{CaDiCaL 2.1.2} \cite{CaDiCaL}: CaDiCaL is a state-of-the-art SAT solver that has achieved high rankings in recent SAT Competitions \cite{CaDiCaLPaper}.
    \item \textbf{CryptoMinisat 5.11.21} \cite{CryptoMinisat}: CryptoMinisat was the winner of the incremental track in the 2020 SAT Competition (no incremental track has been held since 2020).
\end{itemize}

We also considered the IC3 implementations in state-of-the-art model checkers, such as ABC \cite{ABC} and nuXmv 2.1.0 \cite{NUXMVLink}, using their default configurations.

We conducted all configurations using the full benchmark suite from the two most recent Hardware Model Checking Competitions (HWMCC'20 and HWMCC'24), totaling 635 cases (after removing duplicates) in AIGER format \cite{AIGER}. All experiments were performed under identical resource constraints (16GB memory and a 3600s time limit) on an AMD EPYC 7532 processor running at 2.4 GHz. To ensure the accuracy of the results, we verified the results across different checkers to ensure consistency. To ensure reproducibility, we have made the implementations of our experiments available at \cite{Artifact}.

\subsection{Experimental Results}
\begin{figure}[!t]
    \centering
    \includegraphics[width=\textwidth]{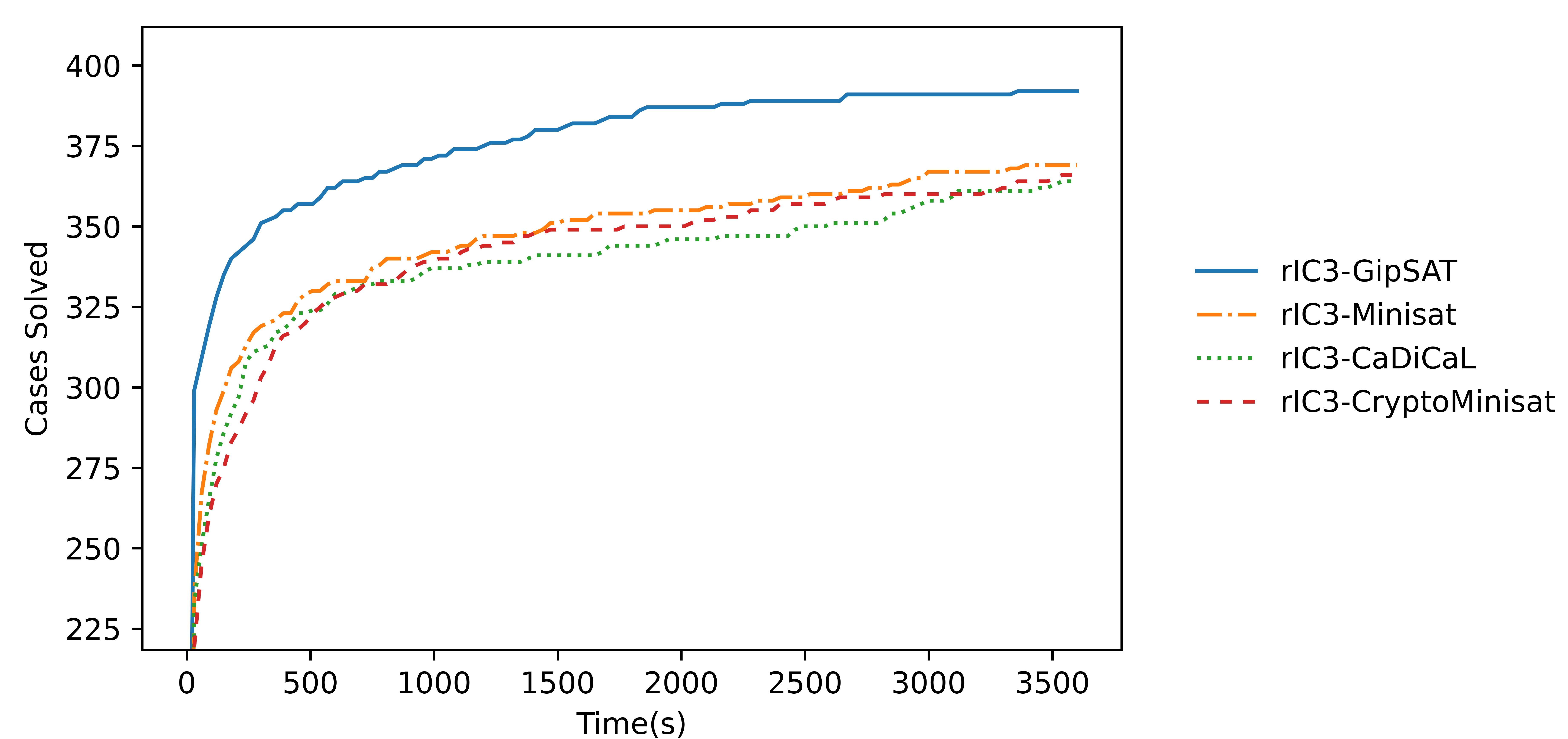}
    \caption{The number of cases solved by different SAT solvers over time.}
    \label{fig:Plot}
\end{figure}

\begin{figure}[!t]
    \centering
    \includegraphics[width=\textwidth]{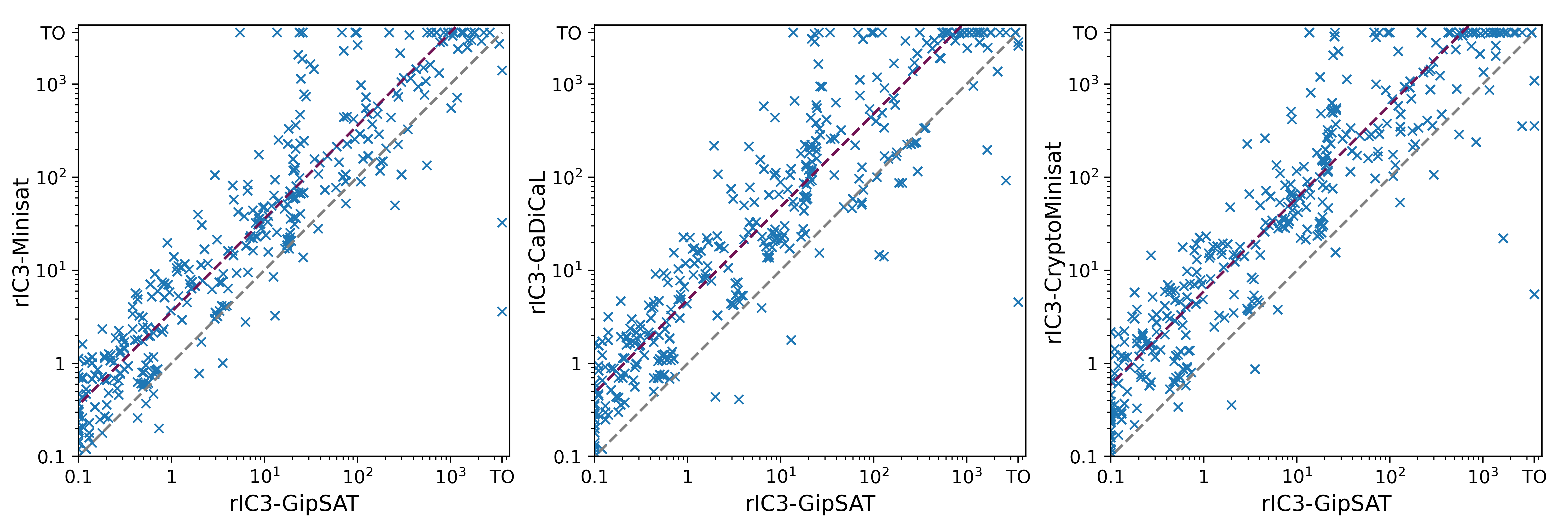}
    \caption{The comparison of solving time(s) between GipSAT and other solvers in rIC3. Points above the gray solid line indicate that GipSAT performs better. The purple dashed line represents the geometric mean of the speedup.}
    \label{fig:Scatter}
\end{figure}

Table \ref{tab:OverallResult} and Fig. \ref{fig:Plot} present the summary of results and the number of cases solved over time by various configurations. The comparison shows that rIC3 performs well relative to state-of-the-art systems like ABC and nuXmv, making it a suitable choice as the experimental foundation. Notably, GipSAT solved $\RICMORE$ more cases than the best-performing regular solver, Minisat, greatly enhancing the capability of the IC3 algorithm. The last column of Table \ref{tab:OverallResult} lists the geometric mean speedup in solving time of GipSAT relative to regular solvers. GipSAT achieves a significant improvement in efficiency, with an average speedup of $\RICSPEEDUP$ times compared to Minisat in rIC3. Fig. \ref{fig:Scatter} shows scatter plots comparing the solving times of GipSAT with those of regular solvers in rIC3. It is clear that, with GipSAT, the solving time for most cases has decreased substantially. This experimental result clearly highlights the effectiveness of GipSAT in improving the efficiency and scalability of the IC3 algorithm.

\begin{table}[!t]
\caption{Total number of cases solved for different configurations. The additional cases solved and the geometric mean of the solving time ratio relative to GipSAT are presented, along with the PAR-2 score for each configuration.}
\centering
\label{tab:OverallResult}
\begin{tabular}{c c @{\hspace{1em}} c @{\hspace{1em}} c @{\hspace{1em}} c @{\hspace{1em}} c @{\hspace{1em}} c}
\hline
Configuration & \textbf{\#}Solved & $\Delta$Solved & \textbf{\#}Safe & \textbf{\#}Unsafe & Avg. ST Ratio & PAR-2\\
\hline
rIC3-GipSAT         &392    &0      &326    &66 &x1.00  &2843.50\\
\hdashline
rIC3-Minisat        &369    &-23    &309    &60 &x3.61  &3140.95\\
rIC3-CryptoMinisat  &364    &-28    &310    &54 &x4.83  &3216.31\\
rIC3-CaDiCaL        &366    &-26    &311    &55 &x5.99  &3189.09\\

\hline
ABC                 &357    &-      &305    &52 &-      &3258.97\\
nuXmv               &353    &-      &301    &52   &-      &3289.72\\
\hline
\end{tabular}
\end{table}

\subsection{Analysis}
\begin{figure}
    \centering
    \includegraphics[width=180pt]{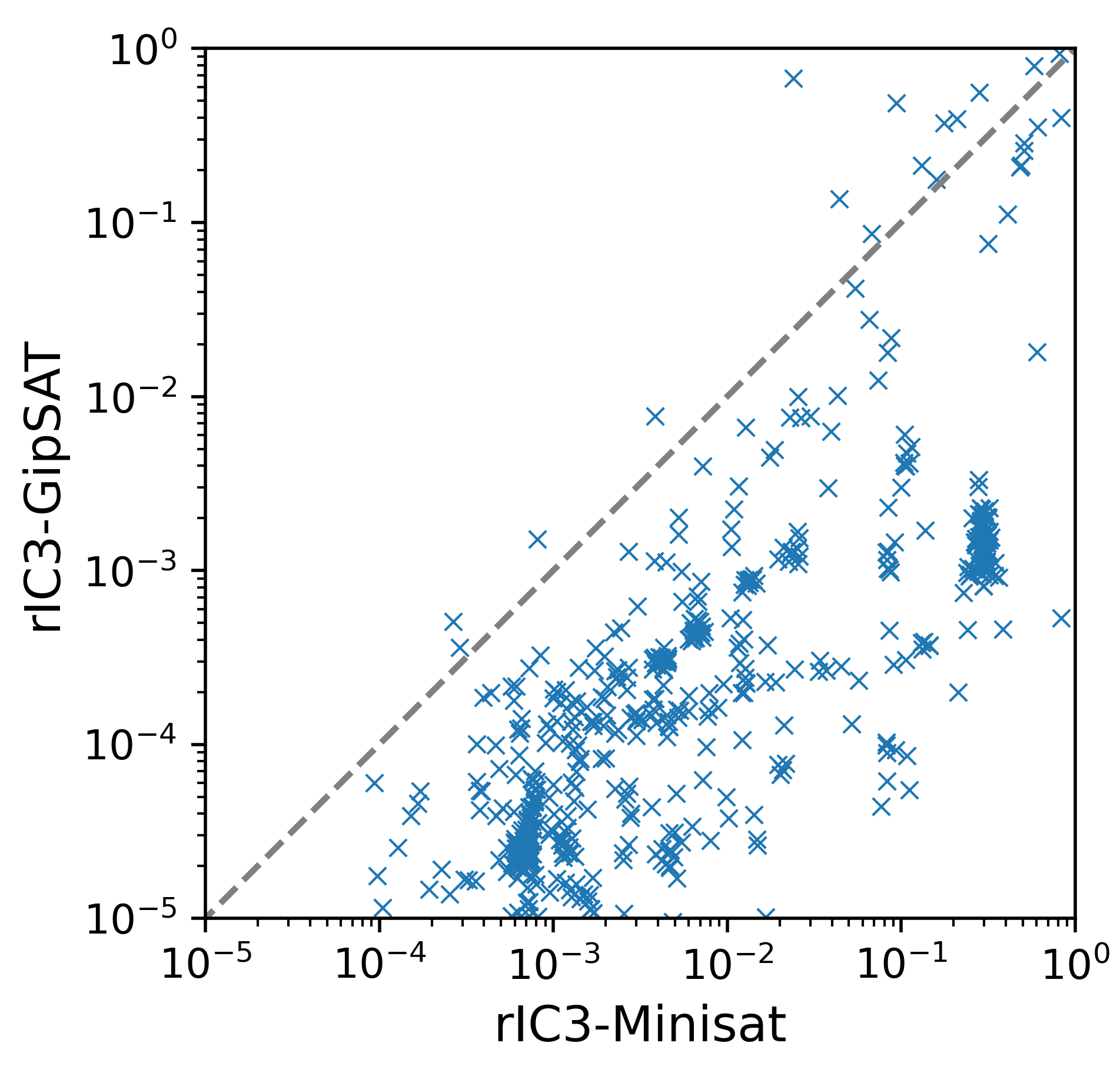}
    \caption{The comparison of average solving time(s) for queries between GipSAT and Minisat in rIC3. Points below the gray diagonal indicate that GipSAT performs better.}
    \label{fig:SatTime}
\end{figure}

\begin{figure}
    \centering
    \includegraphics[width=\textwidth]{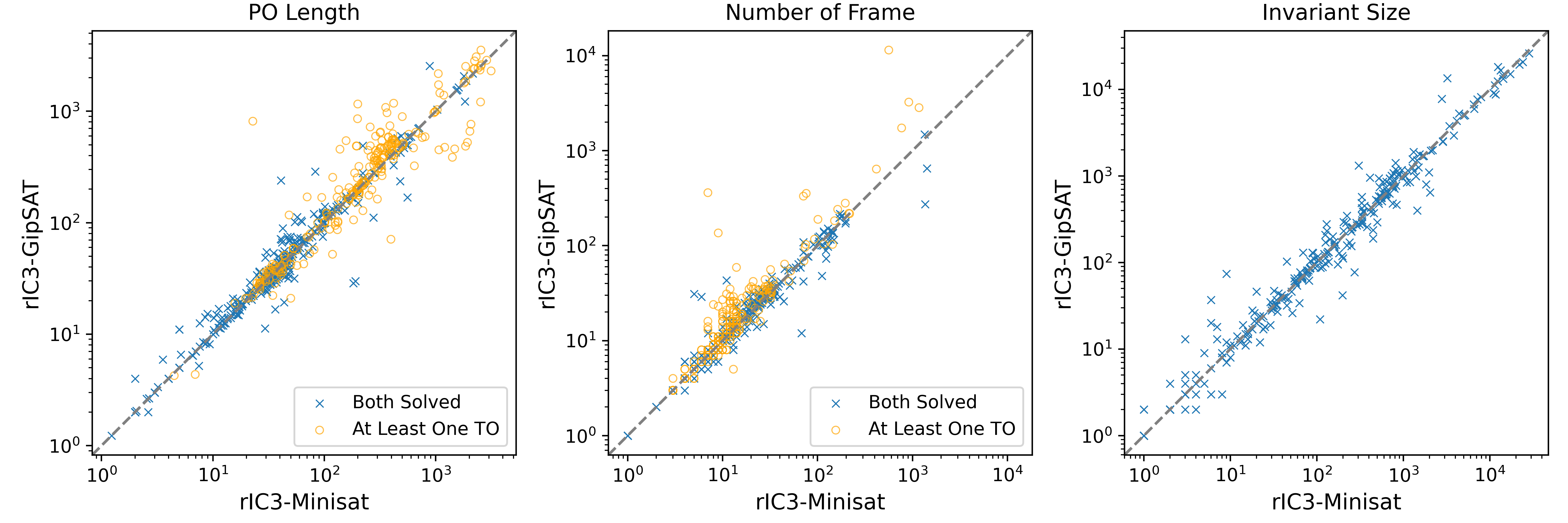}
    \caption{The comparison of proof-obligation length, number of frames, and invariant size (safe instances) between GipSAT and Minisat in rIC3.}
    \label{fig:Inner}
\end{figure}

To better demonstrate the effectiveness of GipSAT, we measured the average solving time of queries using GipSAT and Minisat in rIC3, as described in Fig. \ref{fig:SatTime}. It is evident that, in the majority of cases, GipSAT demonstrates a significant reduction in solving time for queries when compared to Minisat. We also compared the proof-obligation length, number of frames, and invariant size between GipSAT and Minisat, as shown in Fig. \ref{fig:Inner}. For cases solved by both GipSAT and Minisat, these values are largely the same from a statistical perspective, indicating that our approach primarily accelerates SAT solving without significantly affecting the algorithm itself. For cases where GipSAT and Minisat do not both solve the problem, the frame depth of GipSAT is slightly larger. This is because GipSAT's faster solving speed allows the algorithm to reach deeper frames more quickly within the given time limit.

To assess the impact of each optimization, we evaluated rIC3-GipSAT under different optimization combinations:
\begin{itemize}
    \item The \textbf{dm} flag determines whether domain optimization is enabled.
    \item The \textbf{bkt} flag indicates whether buckets are used to maintain VSIDS; if not, a binary heap is used instead.
    \item The \textbf{wr} flag represents whether the optimization described in Section \ref{Sec:WithoutResetting} is applied; otherwise, a new activation variable is created before each solving, and the solver is reset every 1000 activation variables.
\end{itemize}

The results are presented in Table \ref{tab:EachOptimization}, where we also include the results of rIC3-Minisat for comparison. In the table, rIC3-GipSAT-noopt refers to the version with all optimizations disabled. As shown, without any optimizations, rIC3-GipSAT-noopt performs similarly to rIC3-Minisat as GipSAT's CDCL framework is inspired by Minisat. Additionally, it can be observed that enabling optimizations leads to performance improvements, demonstrating the effectiveness of deciding in the domain, using buckets in VSIDS, and eliminating resetting, respectively. For further comparison, we also implemented the method from \cite{TemporaryClause} on GipSAT (rIC3-GipSAT-fmcad21), which performs slightly better than ours. This method sequentially treats each literal in the constraint as an assumption, resulting in support for only one temporary clause, whereas our method allows for multiple temporary clauses.

\section{Discussion}
The IC3 algorithm involves a large number of relatively simple SAT problems, so improving the efficiency of each SAT call is essential for enhancing overall performance. The first optimization significantly boosts performance by reducing the number of decision variables, which leads to substantial improvements in many cases. However, when domain variables make up a large portion of the total variables, the effectiveness of this optimization becomes limited. The second optimization replaces the binary heap with a bucket-based approach to reduce the overhead of the VSIDS heuristic, while the third reuses activation literals to avoid unnecessary resets. Nonetheless, compared to the first optimization, the performance gains from the second and third optimizations are relatively modest. Overall, GipSAT significantly improves the SAT solving efficiency of the IC3 algorithm, enabling more cases to be solved within a limited time. However, it does not alter the IC3 algorithm itself, so cases that are unsolvable due to theoretical limitations remain unsolved.

\begin{table}
\caption{The total number of cases of different conditions. Using rIC3-GipSAT-noopt as the baseline, the additional solved cases and the geometric mean of the solving time improvement compared to the baseline are presented.}
\centering
\label{tab:EachOptimization}
\begin{tabular}{c c @{\hspace{1em}} c @{\hspace{1em}} c @{\hspace{1em}} c}
\hline
Configuration & \textbf{\#}Solved & $\Delta$Solved & Avg. ST Imprv\\
\hline
rIC3-GipSAT-noopt       &370    &0      &x1.00 \\
\hdashline
rIC3-Minisat            &369    &-1     &x0.98  \\
rIC3-GipSAT-dm          &384    &+14    &x2.62  \\
rIC3-GipSAT-bkt         &374    &+4     &x1.24  \\
rIC3-GipSAT-wr          &373    &+3     &x1.13  \\
rIC3-GipSAT-fmcad21     &373    &+3     &x1.21  \\
rIC3-GipSAT-bkt-wr      &379    &+9     &x1.42  \\
rIC3-GipSAT-dm-wr       &386    &+16    &x2.84  \\
rIC3-GipSAT-dm-bkt      &390    &+20    &x3.17  \\
rIC3-GipSAT-dm-bkt-wr   &392    &+22    &x3.52  \\

\hline
\end{tabular}
\end{table}

\section{Related Work}
Since the introduction of IC3 \cite{IC3}, numerous efforts have been dedicated to enhancing its performance. Most of these studies focus on the algorithm layer. For instance, \cite{CTG} attempts to block counterexamples to generalization (CTG) to improve generalization. \cite{AVY} proposes a variant of IC3 that integrates interpolation, while \cite{PDRPROGRESS} introduces under-approximation to expedite bug detection. The algorithm in \cite{CAV23} drops literals that do not appear in any subsumed lemmas from the previous frame, thereby increasing the likelihood of propagating to the next frame. In \cite{PredictingLemma}, the authors aim to predict the outcome before generalization, which could potentially reduce overhead if successful. In \cite{IC3INN}, the authors leverage the internal signals of the circuit to represent invariants more concisely. \cite{IC3KIND} presents a flexible algorithmic framework that integrates IC3 with k-induction. Additionally, \cite{PushToTop} aggressively pushes lemmas to the top by adding may-proof-obligation.

Only a few studies have focused on the SAT solver layer. \cite{IC3SolverManagement} analyzed the results of different SAT solver management strategies at the algorithm layer, including SAT solver allocation, loading, and clean-up. \cite{TemporaryClause} proposed a method that eliminates the need for activation literals in temporary clauses. Instead, it iteratively assumes a literal in the temporary clause before solving, thereby avoiding the need to reset the SAT solver, which means it can only support one temporary clause. Similarly, GipSAT also eliminates the need to reset the SAT solver, but it differs significantly in its approach. GipSAT uses the same activation literal in every query by promptly removing both the temporary clause and its derived learned clauses after each solving, allowing multiple temporary clauses. In Minisat, the \texttt{release\_var} interface can avoid restarts by passing the activation variables. The principle behind it is to first add the activation variable clause and then periodically clean up satisfied clauses (since the occur list is not maintained for performance reasons, it can only be lazily cleaned up). Once all clauses containing released variables are cleared, the activation variables can be reused. Compared to this approach, our method not only avoids restarts but also promptly removes temporary clauses. In SMT, temporary constraints can be supported through \texttt{push} and \texttt{pop}, but more complex data structures are needed to maintain them. 

\section{Conclusion and Future Work}
In this paper, we present GipSAT, a SAT solver specifically designed for the IC3 algorithm. GipSAT reduces the number of variables that need to be decided by precomputing the domain. It achieves constant-time operations by utilizing buckets in VSIDS. Moreover, it avoids resetting the solvers by eliminating the requirement for allocating an activation literal before each solving. The experimental results demonstrate that GipSAT significantly reduces the solving time of SAT queries posed by IC3, consequently leading to a substantial performance improvement compared to state-of-the-art regular SAT solvers. In the future, we will continue to optimize GipSAT and explore the possibility of designing an SMT solver for the word-level IC3 algorithm.

\section*{Acknowledgement}
This work was supported by the Beijing Municipal Natural Science Foundation (Grant No. 4252024), the Foundation of Laboratory for Advanced Computing and Intelligence Engineering (Grant No. 2023-LYJJ-01-013), the Basic Research Projects from the Institute of Software, Chinese Academy of Sciences (Grant No. ISCAS-JCZD-202307) and the National Natural Science Foundation of China (Grant No. 62372438).

\subsubsection*{Disclosure of Interests}
The authors have no competing interests to declare that are relevant to the content of this article.

\bibliographystyle{splncs04}
\bibliography{cites}
\end{document}